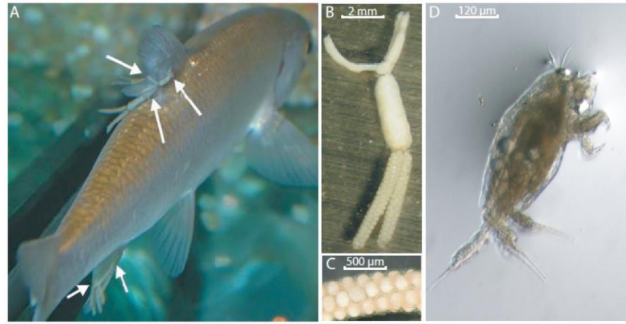

# Molecular approaches reveal weak sibship aggregation and a high dispersal propensity in a non-native fish parasite

Jérôme G. Prunier[1*], Keoni Saint-Pé[1*], Simon Blanchet[1,2], Géraldine Loot[1,2**], Olivier Rey[3**]

[1] Station d'Ecologie Théorique et Expérimentale, UMR 5321, 09 200 Moulis, France.

[2] CNRS, UPS, ENFA, Evolution & Diversité Biologique (EDB) UMR 5174, 118 Route de Narbonne, 31062 Toulouse, Cedex 9, France

[3] CNRS UMR 5244, Interactions Hôtes-Pathogènes-Environnements (IHPE), Université de Perpignan Via Domitia, Perpignan, France

* These two authors contributed equally to this work.

**Authors to whom correspondence should be addressed:

olivier.rey.1@gmail.com

geraldine.loot@univ-tlse3.fr

**Abstract**

Inferring parameters related to the aggregation pattern of parasites and to their dispersal propensity is important for predicting their ecological consequences and evolutionary potential. Nonetheless, it is notoriously difficult to infer these parameters from wildlife parasites given the difficulty in tracking these organisms. Molecular-based inferences constitute a promising approach that has yet rarely been applied in the wild.Here, we combinedseveral population genetic analyses including sibship reconstruction to documentthe genetic structure, patterns of sibship aggregation and the dispersal dynamics of a non-native parasite of fish, the freshwater copepod ectoparasite*Tracheliastespolycolpus*. We collected parasites according to a hierarchical sampling design,withthe sampling of all parasites from all host individualscapturedineight sites spread along an upstream-downstream river gradient. Individual multilocus genotypes were obtained from 14 microsatellite markers, and used to assign parasites to full-sib families and to investigate the genetic structure of *T.polycolpus* among both hosts and sampling sites. The distribution of full-sibs obtainedamong the sampling sites was used to estimate individual dispersal distances within families. Our results showed that *T. polycolpus* sibs tend to be aggregated within sites but not withinhost individuals. We detected important upstream-to-downstream dispersal events of *T.polycolpus*between sites (modal distance: 25.4 km; 95% CI [22.9, 27.7]), becoming scarcer as the geographic distance from their family core location increases. Such a dispersal pattern likely contributes to the strong isolation-by-distance observed at the river scale. We also detected some downstream-to-upstream dispersal events (modal distance: 2.6 km; 95% CI [2.2 – 23.3]) that likely result from movementsof infected hosts.Within each site, the dispersal of free-living infective larvae among hosts likely contributes to increasing genetic diversity on hosts, possibly fostering the evolutionary potential of *T. polycolpus*.

**Keywords**

# 1.    Introduction

Dispersal is a major process influencing ecological and evolutionary dynamics, including the dynamics and persistence of populations, as well as local adaptation and speciation(Clobert et al., 2012; Dieckmann et al., 1999). In parasites, dispersal determinesthe evolution of life-history traits such as their transmission dynamics and their virulence(Barrett et al., 2008; Clayton & Tompkins, 1994; Criscione et al., 2005; Gandon & Michalakis, 2002; Huyse et al., 2005).Parasite dispersalis a complex process thatcan result from the combination of their own movements(when free-living stages exist) and that of their intermediate and/or definitive hosts(e.g., McCoy, 2008; Witsenburg et al., 2015). Over large geographical scales, parasite dispersalis generally considered as being mostly driven by the movements of their hosts/vectors(Blasco-Costa et al., 2012; Feis et al., 2015; Prugnolle et al., 2005; but seeMazé-Guilmo, Blanchet, McCoy, et al., 2016). Yet, dispersal of parasites among hosts also contributes to the overall observed dispersal pattern as soon as a free-living stage occurs, specifically at small spatial scales (e.g, Sire et al., 2001). The individual dispersal among hosts depends on both the intrinsic characteristics of free-living stages, including mobility and survival time, and the environment in which free-living stages are released(Barrett et al., 2008; Boxaspen, 2006; Samsing et al., 2015; Viney & Cable, 2011).

Estimating dispersal of parasites is fundamental to better document and predict their spread, as well as to identify potential source and sink populations of infection(Barrett et al., 2008; Blasco-Costa et al., 2012). From a practical perspective, the above information is useful to design management plans to limit parasite propagation and mitigate their impacts, notably in the case of emergent parasites. The most straightforward -yet challenging- approach to investigate dispersal consists in directly tracking individual movements. Although commonly used for large organisms(Broquet & Petit, 2009; Cayuela et al., 2018; Wikelski et al., 2007), these direct methods are generally unsuited for parasites, notably because of their small size and the difficulty to make them traceable (but seeRieux et al., 2014; Zohdy et al., 2012). Accurate spatio-temporal occurrence data can also be used to indirectly infer dispersal patterns of parasites. This approach iscommonly used in epidemiology to retrace and predict the spatiotemporal dynamics of well monitored parasites and/or pathogens(Ostfeld et al., 2005; Pullan et al., 2012).

The advent ofmolecular approaches hasgreatly contributed to our understanding of parasite dispersal(Blasco-Costa et al., 2012; Giraud, 2004; Mazé-Guilmo, Blanchet, Rey, et al., 2016; Karen D. McCoy et al., 2003; Prugnolle et al., 2005). Molecular tools have mainly been used to infer parasite dispersal indirectly through the use of population genetic structure approaches and/or through phylogenetic analyses(Archie et al., 2009; Lymbery & Thompson, 2012). The examination of parasite population genetic structure at different hierarchical levels of organisation, i.e., within hosts, among hosts from the same site and among sites,is particularly valuableto assess the respective contribution of parasite transmission and host/vector movements to global parasite dispersal(Agola et al., 2009; Bruyndonckx et al., 2009; Dharmarajan et al., 2010; Mazé-Guilmo, Blanchet, Rey, et al., 2016; Mccoy, 2009; Sire et al., 2001). However, these methods often rely on the presence of strong genetic signatures(Faubet&Gaggiotti, 2008; Holderegger&Gugerli, 2012)and may fail to provide accurate estimates of the geographical distances covered by parasites.Alternatively, molecular sibship reconstructioncan be used to assign each parasite to at least one of their parents, their families or their populations of origin based on their multilocus genotypes(Manel et al., 2005). The membership of each parasite to a group, either a population or a family, constitutes individual traceable marks that can be used to explore the distribution of geographical dispersal distances covered by parasites in a way similar to the analyses of "dispersal kernels"(Cayuela et al., 2018; Clobert et al., 2012; Pinsky et al., 2017). Surprisingly, this approach has rarely been used for estimating dispersal parameters of parasite populations(Dubé et al., 2020; Lu et al., 2010).

Here, we empirically tested the value of combining sibship reconstruction to other population genetic tools to assess parasite dispersal and to tease apart the respective contribution of both free-living stages and host-driven dispersal in structuring parasite populations in natural landscapes. We focused on populations of the freshwater ectoparasite copepod *Tracheliastespolycolpus*and its principal local host, the rostrum dace*Leuciscusburdigalensis* (a cyprinid fish),in the ViaurRiver in southwesternFrance. We analysed the distribution of full-sib families and the genetic structure of *T. polycolpus* at different scales, by hierarchically sampling all parasites from all hosts captured within eight sites along the upstream-downstream gradient of the Viaur River. Based on the ecological information available for*T.polycolpus*and its host (see section *Biological model*), we builtseveral non-mutually exclusive predictions. After hatching, the free-living larvae of *T. polycolpus* released into the water column almost instantaneously develop into an infectious stage (Copepodid instar, see Figure 1) allowing a rapid infection of hosts (within a few days; Mazé-Guilmo et al., 2016). Moreover, daces are relatively gregarious and often behave in shoals. We thus expected that parasitesfrom the same clutch wouldmostly infect their natal hosts and/or new hosts from their natal population and would thus mostlyaggregatewithin sites.Alternatively, *T. polycolpus*free-living larvae could passivelydisperse with waterflow (i.e., upstream-to-downstream biased dispersal) over "large" distances until encountering a new host.We thus expectedthat parasites from the same clutch woulddrift,infecting hosts fromdownstream non-natal populations. Finally, because daces are relatively sedentary and their dispersal particularly constrained by several artificial weirs and dams in the ViaurRiver (Blanchet et al., 2010; Clough, 1997; Clough & Beaumont, 1998), we expected that host-drivenupstream-directed dispersal movements of *T. polycolpus*wouldonly occur over short distances.

## 2.    Material and methods

### 2.1. Biological model

Tracheliastespolycolpus is a freshwater ectoparasite copepod that was recently introduced in Western Europe (Rey et al., 2015) and that threatens local populations of daces (*Leuciscus sp.*) and, to a lower extent, several other cyprinid fish species (e.g., chubs, gudgeons, minnows;Loot et al., 2004; Lootvoet et al., 2013). The principal host of *T. polycolpus*is *Leuciscusburdigalensis*(the rostrum dace), with a high prevalence (10% to 90%) when compared to the average prevalence on alternative hosts (1% to 10%;Lootvoet et al., 2013). *Tracheliastespolycolpus*is monoxenous, i.e., it requires a single hostto fulfill its life cycle. The post-embryonic development involves three main stages:nauplius, copepodid and chalimus(Piasecki, 1989). Nauplius is the free-living pre-infective stage. It contains an already formed copepodid inside, whose release can be very quick after hatching (almost immediately or after a few seconds or minutes). A short pre-infective phase is generally considered as an adaptation in parasitic copepods to reach the infective stage as soon as possible, hence maximizing time for infective larvae to encounter and attach on a susceptible host(Piasecki, 1989). Thefree-living infectivecopepodid(Figure 1) displays modest ability to swim and, not adapted to feeding, can live freely for about five days under laboratory conditions(Mazé-Guilmo, Blanchet, Rey, et al., 2016; Piasecki, 1989). Once attached to a host, it transforms into chalimus within five hours. Both sexual dimorphism and mating occur at this stage. Males are dwarf and able to crawl over the host body in search of a female. Females are much larger and attachedto the fins of host, feeding on the mucus and epithelial cells and hence causing lesions and gradually leading to the total destruction of hosts' fins(Loot et al., 2004).The species is monogamous,the female vaginal pore being sealed after fertilization (Piasecki, 1989; sell also Appendix S1). While males usually die very soon after mating (Kabata, 1986), femalescan live up to 89 days(Piasecki, 1989) and producetwo egg sacs each containing up to 165 eggs(Loot et al., 2011).

### 2.2. Sampling design and collection of genetic data

We focused our study on theViaur River, a 169 km-long river located in the Adour-Garonne drainage basin in southwestern France(Figure 2). Eight sites scattered over80.5 km of the whole river upstream-downstream gradient were sampled during the summer 2006(Figure 2; Table 1). Parasites were exclusively sampled on *L.burdigalensis*.Ateach site, daces were sampled using electric-fishing along a 50-200m-long transect using a DEKA 7000, generating 200–500 V with an intensity range of 1 to 3 A. A total of 126 daces was captured and each was anaesthetized using clove oil (30-50 mg/L). The attached parasites to each fin, if any,were counted before being collected using forceps and stored in ethanol for subsequent genetic analyses.All host individuals were then returned alive to their original sampling site.

IndividualDNA extractionswere performed on parasite trunks to avoid any contamination with genetic material from eggs,following a standard salt protocol(Aljanabi & Martinez, 1997). Individual multilocus genotypes were obtained at 16 microsatellite markers (Appendix S2) for each of the 1207 parasites. The 16 microsatellite lociwere co-amplified by PCR in two multiplex batches using the QIAGEN® Multiplex PCR Kit (Qiagen). The two PCR were carried out in a 10 µl final volume containing 5-20 ng of genomic DNA, 5 µl of 2× QIAGEN Multiplex PCR Master Mix, and locus-specific optimised combination of primers (Appendix S2). Both multiplex PCRwere performed in a Mastercycler PCR machine (Eppendorf®) under the following conditions: 15 min at 95 °C followed by 30 cycles of 30s at 94 °C, 90s at 56 °C and 60s at 72°C and finally followed by a 45 min final elongation step at 60°C. The resulting PCR products were separated by electrophoresis on an ABI3730 at the GenoToul(Toulouse France). Allele scoring was performed using GENMAPPER version 4.0.

### 2.3. Preliminary genetic analyses

We first checked for anomalies owed to the genotyping procedure (e.g. large allele dropouts; null alleles) using Microcheckerv2.2(Van Oosterhout et al., 2004). We then tested for linkage disequilibrium between loci and departure from Hardy-Weinberg equilibrium within each sampling site and for each locus using GENEPOP(Raymond & Rousset, 1995), with sequential Bonferroni correction to account for multiple related tests (Rice, 1989). Two markers (TRA12 and TRA66)displayed either strong deficit in heterozygosity, most likely because of the presence of null alleles,or linkage disequilibrium with several other markers. These two loci were therefore discarded from the database in subsequent analyses. Forty individuals were genotyped twice and showed a 100% match in allele scoring at the 14 retained microsatellite markers.

*2.4. Genetic diversity and structure*

Genetic diversity within each of the eight sampling sites was estimated over all loci by computing the unbiased expected heterozygosity($H_e$) using GENETIX(Belkhir et al., 2004),the standardized allelic richness ($A_r$; minimum sample size of 66; Table1)using FSTAT(Goudet, 2001)and the $F_{IS}$index using GENEPOP.Genetic differentiation was assessed by computing the Meirmans'$\varphi_{ST}$(Meirmans, 2006)overall sites and pairwise $\varphi_{ST}$between sitesusing the *mmod* R-package (Winter, 2012). The effective population size Ne of*T. polycolpus*within the Viaur River (all individuals combined) was estimated using NeEstimatorv.2.1(Do et al., 2014)based on a linkage disequilibrium methodand setting the lowest allele frequency to 5%,considering monogamous mating andusing 95% confidence intervals based on Jackknife resampling. We expected Ne to be small, since metazoan parasites generally have smaller effective population sizes than free-living species (Criscione & Blouin, 2005).

We thenexplored how the genetic diversity of *T. polycolpus*wasgenetically and spatially structured among sampling sites along the ViaurRiver using threeindependent approaches. First, we tested whether the global spatial pattern of genetic differentiation between sites along the ViaurRiver followed a pattern of isolation-by-distance. To do so, we performed a mantel test between matrices of pairwise measures of genetic differentiation and geographical riparian distances (i.e., geographical distances along the water course; Blanchet et al., 2010)between sites using the R-package *vegan*(Oksanen et al., 2020). The Mantel correlation *r* was computed and the associated *P*-valuewas calculated using 10,000 random permutations.Additionally, we performed a non-directional Mantel correlogram(Borcard& Legendre, 2012; Smouse&Peakall, 1999)using the R-package *ecodist*(Goslee& Urban, 2007)with one-sided Mantel tests with 1,000 permutationsand geographical riparian distance classes defined every 10km (up to 80 km). Secondly, we performeda discriminant analysis of principal components(dAPC) using the R-package *adegenet* (Jombart et al., 2010) for a visual assessment of between-sites differentiation. Finally, we performed an analysis of molecular variance (AMOVA) using ARLEQUIN V.3.5 (Excoffier & Lischer, 2010) to measure the amount ofoverall genetic variance of *T. Polycolpus*explained by each of the three hierarchical structure levels considered within the ViaurRiver: (i) within hosts, (ii) among hosts within sites and (iii) among sites.

## 2.5. Reconstruction of full-sib families

Full-sibs families of *T. polycolpus* were reconstructed using the full-likelihood approach implemented in COLONY2.0(Jones & Wang, 2010)based on the 1207 individual multilocus genotypes.Briefly, COLONY 2.0implements full-pedigree likelihood methods, i.e., with likelihood considered over the entire pedigree, to infer sibship among individuals. We assumed that both sexes are monogamousand we allowed for possible inbreeding.All individuals were considered as offspring in COLONY 2.0 and we defined no *a priori* candidate parental genotypes (neither males nor females). Allele frequencies were directly determined from the genetic dataset using COLONY version 2.0. Only the full-sib families with associated inclusion probability higher than 95 % were retained for further analyses.

## 2.6. Distribution of full-sib families

We first assessed whether full-sib individualswere rather clumped within the same site or randomly distributed across sites.To do so, we built two binary matrices that respectively included (i) the membership status of each pair of individuals to the same family (i.e., 1: parasites are full-sibs, 0: parasites are not full-sibs; hereafter called the sibship matrix) and (ii) the membership status of each pair of individuals to the same site (i.e., 1: parasites share the same site, 0: parasites come from different sites; hereafter called the site matrix). Based on our observed dataset, we computed the proportion of full-sib pairs sharing the same site (i.e., pairs of individuals displaying values of 1 in the two matrices) given the total number of full-sib pairs over the river(i.e., pairs of individuals that displayed value of 1 in the sibship matrix). This observed proportion was compared to a series ofexpected proportions under the null hypothesis of a random distribution of full-sib pairs among sites, using 10,000 random permutations of the site matrix to compute the probability of correctly rejecting the null hypothesis(Legendre & Legendre, 1998).

Similarly, we assessed whether full-sib individuals were rather clumped on the same host or randomly distributed across hosts.Because hosts were not distributed homogeneously among the eight sampling sites, we considered each site independently. For each site, we first built two binary matrices that respectively included (i) the membership status of each pair of individuals to the same family (sibship matrix) and (ii) the membership status of each pair of individuals to the same host (i.e., 1: parasites are on the same host, 0: parasites areon different hosts; hereafter called the host matrix).We then computed the proportion of full-sib pairs sharing the same host (i.e., pairs of individuals displaying values of 1 in the two matrices) given the total number of full-sib pairs within the considered siteand compared this observed proportion to a series of expected proportions under the null hypothesis of a random distribution of full-sib pairs among hosts, using 10,000 random permutations of the host matrix.

### 2.7. Estimation of T. polycolpus dispersal

To investigate the dispersalof *T. polycolpus* along the ViaurRiver, we focused on a subset of full-sib families including at least 5 full-sibs (N = 94 families). For each of these 94 families, we first determined a "family core location"as the mode of the kernel distribution of the geographical distance of each family member to the river sourceusing theR-package *stats*(R core team; 2014). Next, we computedfor each of the 94 families(i)a "downstream maximal dispersal distance" estimated as the difference between the estimated "family core location"and the distance of the most downstream family member to the river source, and (ii)an "upstreammaximal dispersal distance" estimated as the absolute value of the difference between the estimated "family core location"and the distanceof the most upstream family member to the river source. We then calculated the modes of the distributions of both the downstream and the upstream maximal dispersal distances across the 94 families. These modes provide a proxy of the most common maximal downstream and upstream distances covered by *T. polycolpus*from the family core location.We computed 95%confidence intervalsabout theseupstream and downstream distance modes using10,000bootstrap replicates. Finally, we testedwhether the upstream and downstream maximal dispersal distances from the family core location were significantly different using a non-parametric Wilcoxon test implemented in the R-package *stats* (R core team; 2014).

## 3. Results

### 3.1. Genetic diversity and structure

From 126 captured daces, 114 were infected by *T. polycolpus*(parasite prevalence over all sites of 90.5%) with a parasite load of 13.4 ± 13.2 (mean ± SD). A total of 1207 parasites were sampled from infectedhosts. Over all sampling sites, $H_e$ was 0.52 ±0.01(mean ± SD), *Ar*ranged from 3.43 to 4.05 and $F_{IS}$ ranged from -0.02 to 0.01(Table 1). The effective population size Ne of *T. polycolpus*at the river scale was537.6 (95% CI [334.2,885.1]).The mean genetic differentiation estimated overall sites and overall loci was $\varphi_{ST}$= 0.08and pairwise $\varphi_{ST}$values between sites ranged from 0to 0.21, suggesting weak to moderate genetic structure in the Viaur River.We found however a strong and significant correlation between pairwise $\varphi_{ST}$and pairwise riparian distances between sites ($r$= 0.90; P-value< 0.001) as expected under an isolation-by-distance pattern (Figure 3A).Additionally, the non-directionalMantel correlogram indicated that parasites from sites distant by less than 20 km tend to be more genetically similar than expected by chance (Figure 3B).These results are in accordance with the high overlap observed between sites within the retained dAPC parameter space (two first components, together explaining 92.3% of variance) and the slight upstream-to-downstream gradient along the first component (Figure 3C).

According to the AMOVA analysis,most of the genetic variation in*T. polycolpus* in the ViaurRiver was actually observed within individual hosts (i.e., 97.9%; $\Phi_{ST}$ = 0.021;*p-value*< 0.01; Table 2). The "among site" level explaineda weak (yet significant) amount of total genetic variation (2.17%; $\Phi_{CT}$ = 0.022;*p-value*< 0.01; Table 2),whereas no partition of the total genetic variation was attributed to the "among hosts within sites" level ($\Phi_{SC}$ = 0;*p-value* = 0.52; Table 2).

### 3.2. Reconstruction and distribution of full-sibs families

Overall, 1075out of the 1207 genotyped parasites were assigned to 160 full-sib families with a probability higher than 95%.On average, reconstructed full-sib families were composed of 6.8 individuals (ranging from 1 to 35; AppendixS3).

We found that 21.0% of the5450 full-sib pairs reconstructed over the ViaurRiver belonged to the same site (Figure 4A). This proportion, although moderate, was significantly higher than the expected theoretical proportion (14.3%) under the null hypothesis (i.e., pairs of full-sibs are distributed randomly across the eight sampling sites; $\chi^2$ = 200.75, d.f. = 1,P-value < 0.001). Moreover, the proportion of full-sib pairs belonging to the same site and infecting the same host differed slightly (but significantly) between sites, andranged from 5.7 to 23.6 % (Figure 4B). Yet, none of these localproportions significantly differed from the expected theoretical proportions under the null hypothesis (i.e., pairs of full-sibs are distributed randomly over the sampled hosts within each site; AppendixS4).

### 3.3. Estimation ofT. polycolpus dispersal

The family core location estimated for each reconstructed full-sib families with more than five full-sibs ranged from 48.6 to 127.2 km from the river source (mean = 68.3 km; Appendix S5).The downstream maximal dispersal distance from the familycore location ranged from 0 to 77.9 (mode = 25.4 km,95% CI[22.9, 27.7]; Figure 5).The upstreammaximal dispersal distancefrom thefamilycore location was significantly lower than the downstream distance (P-value < 0.01) and ranged from 0 to 78.6 km (mode = 2.6 km, 95% CI[2.2,23.3]; Figure 5).

## 4. Discussion

By combining sibship reconstruction with more classical population genetic tools, we were able to estimate various dispersal parameters in the parasite *T. polycolpus*. Hereafter we will discuss the dispersal dynamics of *T. polycolpus* among hosts and among sites at a river scale. We will also discuss the relative contribution of the passive copepodid dispersal and of the host-driven chalimus dispersal in shaping the genetic structure of populations as well as the possible evolutionary outcomes of such dispersal dynamics for both parasite and host populations.

We found that most of the full-sib family members of *T. polycolpus* do not infect the same host -and hence not their natal hosts- as a clump but are rather scattered over several host individuals. Consequently, at each generation, the dispersal of free-living copepodids among hosts probably contributes to the genetic mixing of unrelated adult breeders within hosts. Accordingly, the AMOVA revealed that most of the genetic variability of *T. polycolpus* along the river occurs within hosts (Table 2). The research of a sexual partner by males occurring once on the host, this dispersal strategy may contribute to limit the probability of mating between related individuals (random mating) and to minimize the possible detrimental effects resulting from inbreeding depression. Theoretical models predict that multi-infection of hosts by parasites from distinct strains can increase parasite virulence (Buckling & Brockhurst, 2008; López-Villavicencio et al., 2011). Local freshwater fish species and specifically daces from the Viaur River may thus suffer from virulent *T. polycolpus* variants. This is in line with a previous studies showing that the pathogenic effects induced by *T. Polycolpus* in the Viaur River is severe (Loot et al., 2004), and, combined with high prevalence, that they might have been responsible for the serious demographic decline of daces locally observed over the last decade (Mathieu-Bégné et al., 2019).

At the site level, a substantial (and significant) fraction of the overall reconstructed full-sib pairs was found to be 'aggregated' within sites (i.e., 21.0 %). This pattern of within site 'aggregation' strongly suggests that some recently hatched *T. polycolpus* infective larvae are able to persist on their natal site by infecting susceptible hosts in the close neighbourhood of their natal hosts. Two non-exclusive ecological factors may account for such a pattern. First, the very short lifetime of the non-infective naupliusstage of *T. polycolpus* (Piasecki, 1989) is likely to facilitate their attachment to host individuals neighbouring their natal hosts as soon as they are released into the water column. Second, daces are gregarious and commonly form shoals(Keith et al., 2011). Local congregations and frequent social interactions between dacehosts may improve host-to-host transmission of parasiteswithin sites(Johnson et al., 2011), all the more so as *T. polycolpus* has recently been shown to preferentially occur at very specific microhabitats that maximize encounter rate and create hotspots of infection (Mathieu-Bégné et al., 2020). Parasite transmission between neighbouring hosts inhabiting the same location is expected to homogenize the genetic variation among hosts at the site level (e.g., Bruyndonckx et al., 2009). Accordingly, the AMOVA revealed that the "among hosts within sites" level did not contribute significantly to the overall genetic variation of *T. polycolpus* in the ViaurRiver.

At the river level, and despite the significant within-site 'aggregation' pattern of full-sibs, the overall genetic structure was weak and reconstructed families were generally disseminated over several sites, indicating successful dispersal events. The overall genetic structure was characterized bya strong isolation-by-distance pattern (Figure 3A) that suggests, according to Hutchison and Templeton (1999), that populations of *T. polycolpus*were at migration-drift equilibrium. This isolation-by-distance pattern also conforms to the results obtained from the AMOVA, which indicates that a significant fraction of the overall genetic variability of *T. polycolpus* along the river occurs among sites.With dispersal among hosts facilitating random mating and dispersal among sites resulting ingene flow, the hierarchical dispersal strategy of *T. polycolpus*probably contributes to maintaining high genetic diversity (high expectedheterozygosity*He* and low $F_{IS}$ values; Table 1) despite limited effective population sizes (Criscione & Blouin, 2005)and may explain the reported invasion success of *T. polycolpus*(Mathieu-Bégné et al., 2019; Mathieu-Bégné et al., 2020; Rey et al., 2015).

Determining the respective contribution of free-living copepodid dispersal and host-driven chalimusdispersal is challenging. Yet, several lines of evidence may help disentangling these two modes of dispersal. As for most riverine free-living organisms with low dispersal ability, copepodids are expected to drift passively downstream their hatching sites due to the unidirectional water flow(Paz-Vinas & Blanchet, 2015). We accordingly detected an upstream-to-downstream dispersal bias from the estimated core location of *T. polycolpus*families, with the majority of downstream dispersal events occurring over the first 25.4 km. It is noteworthy that this direct estimate of downstream dispersal distance is highly congruent with the Mantel correlogram (Figure 3b), with demes becoming genetically differentiated as soon as they are distant from more than ~20 km. Host-driven dispersal of fixed adult parasites is also likely to contribute to the overall dispersal of *T. polycolpus* along the river. However, daces are relatively sedentary, spending extended periods in a single site before moving towards surrounding sites within a mean radius of two kilometres and up to ten kilometres over the year (Clough, 1997; Clough & Beaumont, 1998). Moreover, dispersal of daces in the Viaur River is highly limited given the important number of obstacles (weirs and dams) that scatter the river (~1 obstacle every 2-3 kilometres in average; Blanchet et al., 2010). This suggests that host-driven dispersal of *T. polycolpus* either downstream or upstream may regularly occur, but may be limited over short geographical distances. Thus, we argue that long upstream-to-downstream dispersal events of *T. polycolpus* (twice the distance covered by their hosts annually; Blanchet et al., 2010; Clough, 1997) likely result from the drift of free-living infectious larvae with waterflow. At smaller geographical scale, dispersal of *T. polycolpus* may be driven by the combination of both free-living and host-driven movements. Interestingly we also detected some downstream-to-upstream dispersal events that mostly occur over short geographical distances (i.e., 2.6 km; Figure 5). The swimming ability of copepodids is clearly insufficient to overcome the water flow of the Viaur River (Piasecki, 1989). Thus, the downstream-to-upstream dispersal of *T. polycolpus* detected testifies the frequent although spatially constrained host-driven movements from downstream to upstream sites once the infective larvae are fixed to their host.

Overall, these conclusions about *T. polycolpus* dispersal strategy are based on the use of an original methodological framework that was made possible by the specific life history traits of both the considered parasite and its host: *T. polycolpus* is a strictly aquatic and monoxenous parasite (i.e., a single host is required to fulfill its life cycle) that is mostly found on *L. burdigalensis* in the studied system. The latter showing both small population sizes and spatially limited movements in the studied system (Mathieu-Bégné et al., 2019), we probably sampled a representative proportion of both hosts and parasites at each sampling site (total sample size twice as high as estimated total effective population size). Furthermore, the monogamous mating system of the parasite strongly facilitated the reconstruction of family groups. We acknowledge that this approach might be more difficult to implement in other host-parasite systems, such as in terrestrial habitats or with species showing more complex life history traits.

## 5.    Conclusion

Documenting the hierarchical genetic structure and quantifying the dispersal of parasites is crucial to better understand their evolutionary potentialand dynamics. By combining various population genetic tools including sibship reconstruction, we found that *T. polycolpus* sibs tend to be aggregated within sites but not within hosts. This pattern may contribute to maintain highgenetic variation on each host through random mating, with possible positive evolutionary outcomes in terms of individual fitness and/or parasitic virulence. We alsodeciphered the relative importance of free-living dispersal of *T. polycolpus*and host-driven dispersal of fixed adults along the river. Our results suggest that *T. polycolpus* displays a substantial ability to dispersethroughout its lifetime, through passive downstream dispersalat the copepodid stage andthrough host-drivenupstream dispersal at the chalimus stage. This hierarchical dispersal strategymay contribute to maintaining high genetic diversity despitelimited effective population sizes and is probably one of the various traits that may explain the invasion success of *T. polycolpus*since its recent introduction within the ViaurRiver and most likely overall French watersheds(Mathieu-Bégné et al., 2020; Rey et al., 2015).

## Acknowledgments

This work is part of the project INCLIMPAR (ANR-11-JSV7-0010) supported by a grant from the Agence National de la Recherche awarded to GL. OR is also grateful to the "Region Midi-Pyrénées" for financial support. The genetic data were generated at the molecular genetic technical facilities of the Genopole Toulouse (Toulouse, France). This work has been done in two research units (SETE & EDB) that are part of the "Laboratoired'Excellence" (LABEX) entitled TULIP (ANR-10-LABX-41).

## Authors' Contributions

JGP:Formal analysis (Equal), Investigation (Equal), Methodology (Equal), Writing-original draft (Equal), Writing-review & editing (Equal). KSP: Formal analysis (Equal), Investigation (Equal), Methodology (Equal), Writing-original draft (Equal), Writing-review & editing (Equal). SB: Conceptualization (Supporting), Methodology (Supporting), Project administration (Supporting), Supervision (Equal), Writing-review & editing (Equal). GL: Conceptualization (Equal), Funding acquisition (Lead), Methodology (Supporting), Project administration (Supporting), Supervision (Equal), Writing-review & editing (Equal). OR: Conceptualization (Equal), Funding acquisition (Supporting), Investigation (Equal), Methodology (Equal), Project administration (Lead), Supervision (Equal), Writing-original draft (Equal), Writing-review & editing (Equal).

## Ethics Statement

Fieldwork was conducted with adequate administrative permits for electrofishing (Permit #2005-34-4 delivered by the "Direction Départemental de l'Aveyron") and fish were treated in accordance with the French Law (Use of live animals for scientific purposes; Articles R214-87 to R214-137 of the rural code).

## Data Accessibility Statement

Microsatellite data are available on Figshare: 10.6084/m9.figshare.14038901.

## Conflict of Interest

The authors declare no conflict of interest.

**Table 1:** Sampling sites of *T. polycolpus* over the River Viaur and genetic diversity estimated across loci at each sampling site or averaged across sites (ALL). $A_r$: Mean standardized allelic richness; $H_e$: expected heterozygosity.

| Sampling site | Locality | Distance from the source (km) | $N_{Hosts}$ | $N_{Parasites}$ | $A_r$ | $H_e$ | $F_{IS}$ |
|---|---|---|---|---|---|---|---|
| V01 | Bannes | 48.61 | 14 | 231 | 3.93 | 0.53 | 0.012 |
| V02 | Capelle | 52.14 | 18 | 257 | 3.95 | 0.53 | -0.014 |
| V03 | Fuel | 67.23 | 12 | 108 | 4.05 | 0.52 | -0.029 |
| V04 | Serres | 69.44 | 15 | 100 | 3.87 | 0.51 | -0.012 |
| V05 | Albinet | 75.15 | 18 | 200 | 3.76 | 0.52 | -0.016 |
| V06 | Navech | 93.77 | 17 | 136 | 3.73 | 0.50 | -0.025 |
| V07 | Just | 99.97 | 12 | 109 | 3.73 | 0.52 | -0.007 |
| V08 | Calquiere | 129.13 | 8 | 66 | 3.43 | 0.53 | -0.015 |
| ALL | | | 14.25 | 150.88 | 3.81 | 0.52 | -0.01 |

**Table 2:** Results of the Analysis of Molecular Variance (AMOVA). D.f.: degrees of freedom.

| Source of variation | D.f. | Sum of squares | Variance components | % of variation |
|---|---|---|---|---|
| Among sites | 7 | 191.93 | 0.08 | 2.17 |
| Among hosts within sites | 106 | 382.14 | -0.002 | -0.07 |
| Among individuals within hosts | 2300 | 8417.65 | 3.66 | 97.90 |

**Figure legends**

**Figure 1.**Pictures of *Tracheliastespolycolpus*atdifferent stages. A: parasiticadultfemales at chalimus stage (indicated by white arrows) attached to a host (***Leuciscusburdigalensis***). B: mature parasiticadultfemalecarryingtwoeggs sacs. C: eggs of *T. polycolpus*enclosedwithin a maternalegg sac. D: recentlyhatchedfree-living copepodidlarvaeready to infect a new host.

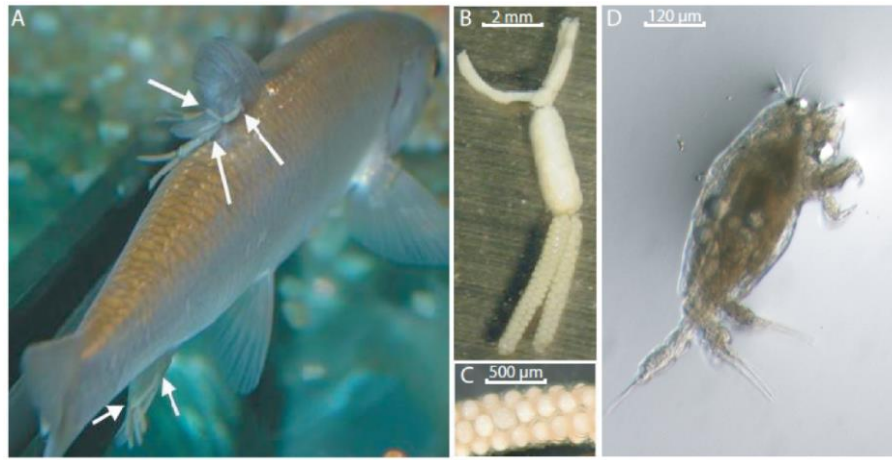

**Figure 2.**Localisation of the eight sampling sites along the River Viaur in France. Tributaries are in light grey.

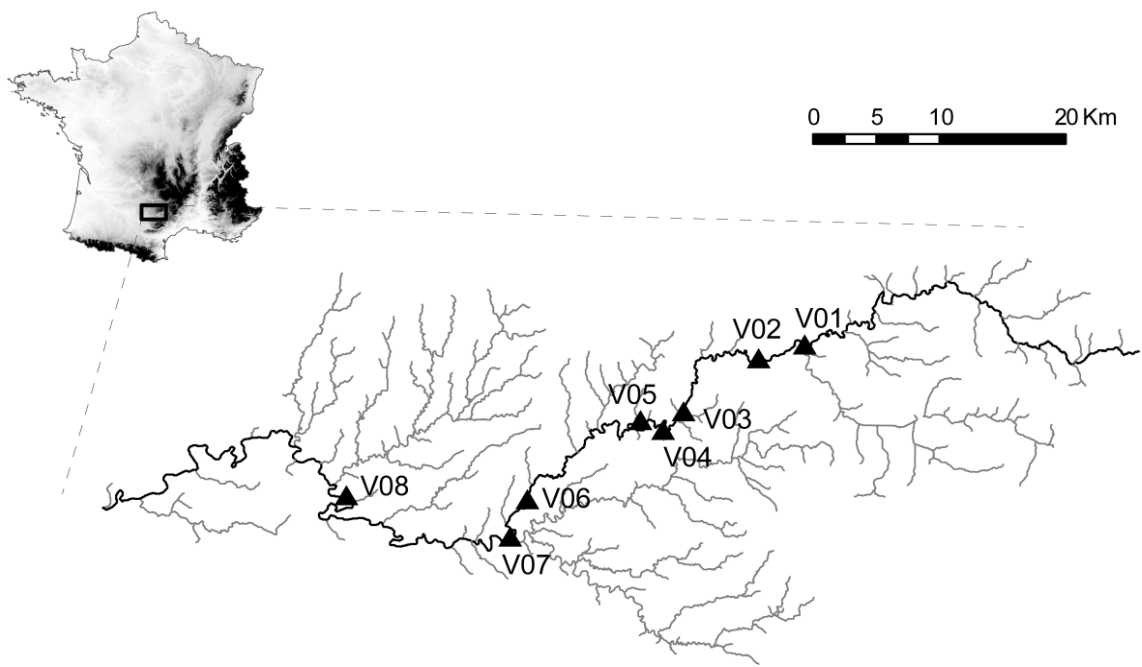

**Figure 3.** A: Scatterplot and best fit linear trend of the Mantel test relating pairwise estimates of genetic differentiation $\varphi_{ST}$ and pairwise riparian geographical distances between sites. B:Scatterplot of the non-directional Mantel correlogram, representing Mantel correlation values (r) obtained between pairwise estimates of genetic differentiation $\varphi_{ST}$ and pairwise riparian geographical distances between sites, with riparian distances classes defined every ten kilometres. Grey points stand for significant (or very close to significance) p-values. Error barsboundthe95%confidenceintervalabout *r* values as determined bybootstrapresampling.. C: Scatterplot of individuals along the two first components of the dAPC and barplot of eigenvalues; each color (points and ellipse) of the scatter plot represent a sampling site.

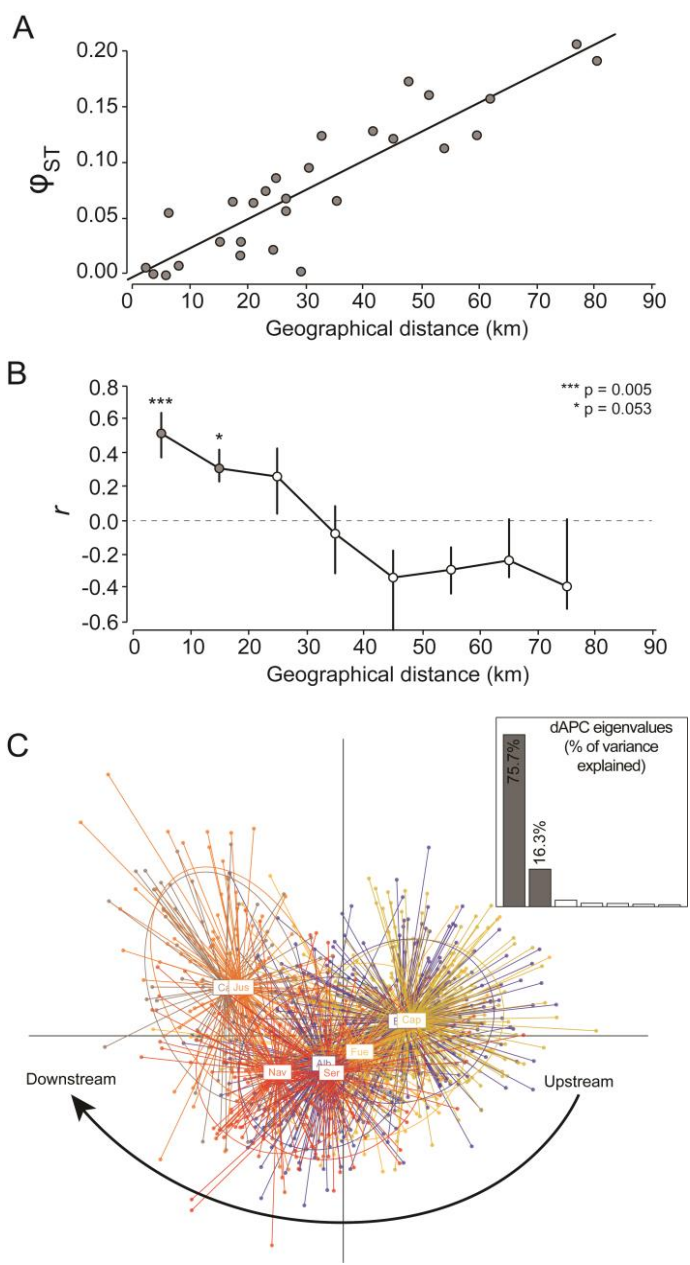

**Figure 4.** Percentage of the reconstructed full-sib pairs sharing the same host (black boxes), the same site (grey boxes) and different sites (white boxes) along the whole river (A) and within each sampling site (B). The lower case letters in B indicate sites that do not differ statistically in the percentage of full-sib pairs sharing the same host.

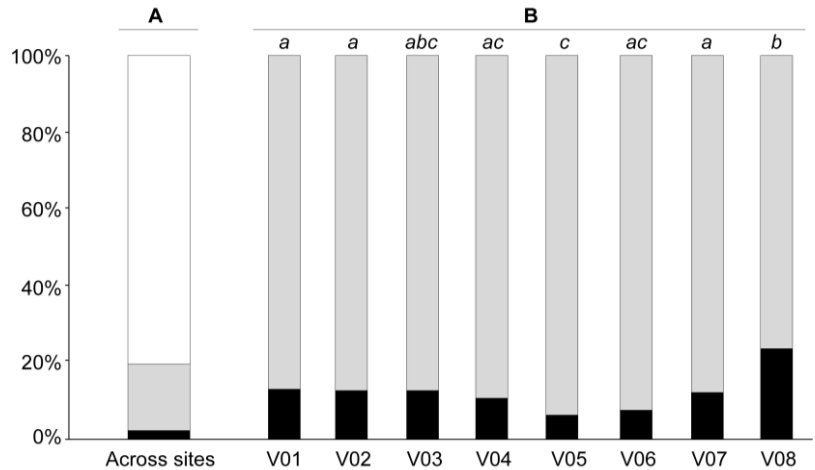

**Figure 5.** Distribution of the upstream and downstream maximal distances covered by individuals from the core location of their family. Only families with more than five full-sibs were considered (n = 94). The modes of the upstream and downstream maximal distances distributions are indicated by dotted lines. The 95% confidence intervals around the upstream and downstream distance modes are highlighted in shaded grey.

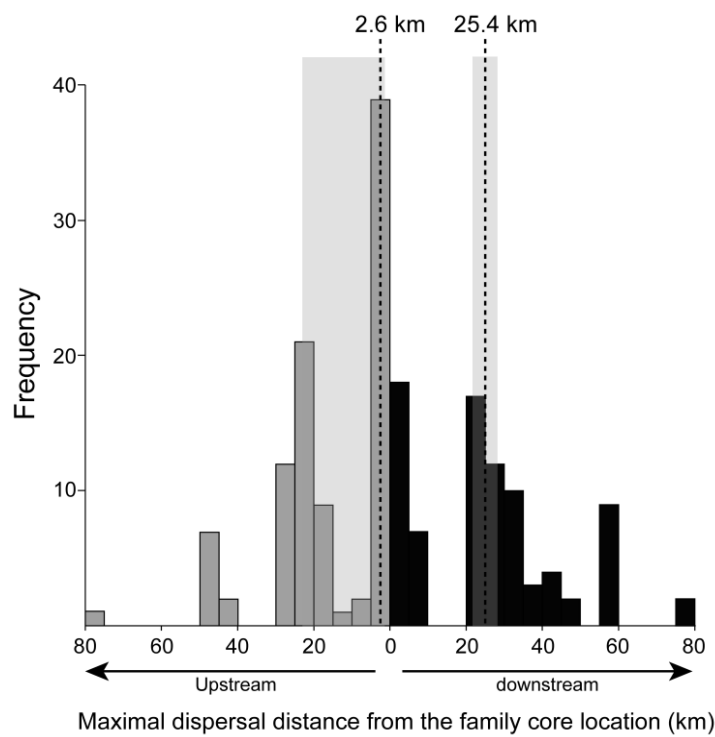

# APPENDICES

**Appendix S1.** As a part of preliminary work for a previous study (Loot et al., 2011), we genotyped 20 eggs (5 eggs per egg sac per female for 2 females) at several polymorphic microsatellite loci and found a maximum of 2 alleles per locus across eggs from the same female, clearly indicating that all eggs originate from the same unique father (unpublished data).

**Appendix S2.** Main characteristics of the 16 microsatellite markers used in

*Tracheliastespolycolpus* and composition of the two PCR multiplexes. Discarded loci are in italics. He:

expected heterozygosity; Ho: observed heterozygosity; Na: number of alleles.

| PRIMER NAME | MULTIPLEX (and DYE) | PCR PRODUCT SIZE (min-max) | PRIMER LEFT SEQUENCE | PRIMER RIGHT SEQUENCE | MOTIF | He | Ho | Na |
|---|---|---|---|---|---|---|---|---|
| TRA5 | B (FAM) | 151-159 | CAGTGGGCAACAAAGGAAAT | AGTTTGGGGAAAACTTGGCT | ATTG | 0.130 | 0.131 | 4 |
| TRA6 | B (HEX) | 193-197 | AAAGGAAAGGCATTTGACCC | GGATGTGTCGAAAAAGCGAT | CCTT | 0.485 | 0.504 | 2 |
| *TRA12* | *A (FAM)* | *164-184* | *CTCAATCAACAAACAAAAGTTTC* | *GAAAAAGCTCAGACAAAAGTATGC* | *TATC* | *0.434* | *0.439* | *6* |
| TRA20 | B (FAM) | 195-197 | CAGGATGAGGATGAATTAGCG | CACGTTGTTTTTGATTCGGA | AG | 0.362 | 0.367 | 2 |
| TRA33 | B (ATTO550) | 188-196 | CAGTCGAAAGTGCGCAATAA | ATTTTTGAGGCGACTACGGA | AG | 0.669 | 0.678 | 5 |
| TRA42 | A (FAM) | 264-280 | CACAAAGTGCATGGAAATGG | GCGTCTCTGAAGGTGATTACG | GA | 0.640 | 0.663 | 5 |
| TRA44 | A (HEX) | 275-279 | CCTGAATTCGAATGAAAACCA | GGAGAATGTAATCGGAAAATCC | GA | 0.638 | 0.639 | 3 |
| TRA49 | B (HEX) | 288-294 | AGGACTTTCGGAGCTGATGA | CACACAGACAGACACAGTCACAA | GA | 0.511 | 0.505 | 5 |
| TRA53 | B (ATTO550) | 314-322 | CCCACAAAAATGCTTTGGTT | CATCTTTGTAAGACGAGTTTGCTG | TC | 0.572 | 0.566 | 5 |
| TRA58 | A (FAM) | 316-330 | TATCGCTGAGAAAGGCACTG | GACGTTTCTACCGGCACTTC | GA | 0.513 | 0.507 | 5 |
| TRA59 | A (HEX) | 330-336 | GGCAGCATCAAATTGTTTGT | CTCGAGACTTTACGGCCATC | TC | 0.479 | 0.497 | 4 |
| TRA62 | B (FAM) | 344-366 | CGGGAAGAGGACGAAAAGA | TCATTTCCGATGATGGCATA | TC | 0.691 | 0.686 | 6 |
| *TRA66* | *B (HEX)* | *351-361* | *TCCTGTATGCCCAAAACACA* | *TCATCAAATAATAAAATCCTTCTTTTCTC* | *AG* | *0.393* | *0.210* | *4* |
| TRA76 | A (ATTO550) | 115-139 | CAGCAACTGTAAGATATTAGCAAC | ACTCGCCGCTAAACACAAG | TA | 0.681 | 0.698 | 12 |
| TRA81 | A (FAM) | 111-115 | TGTCCATCTCCTTGAGTGGC | AGGCCTCGTTCCTTCGTATT | AG | 0.417 | 0.423 | 3 |
| TRA90 | A (ATTO550) | 189-199 | AGATGTCAAACCCTGGGATG | TTCCATAACCCAACAGGGAC | TC | 0.473 | 0.486 | 5 |

**Appendix S3.** Distribution of the number of full-sib membersobtainedin the 160 reconstructed full-sib familiesof *T. polycolpus*.

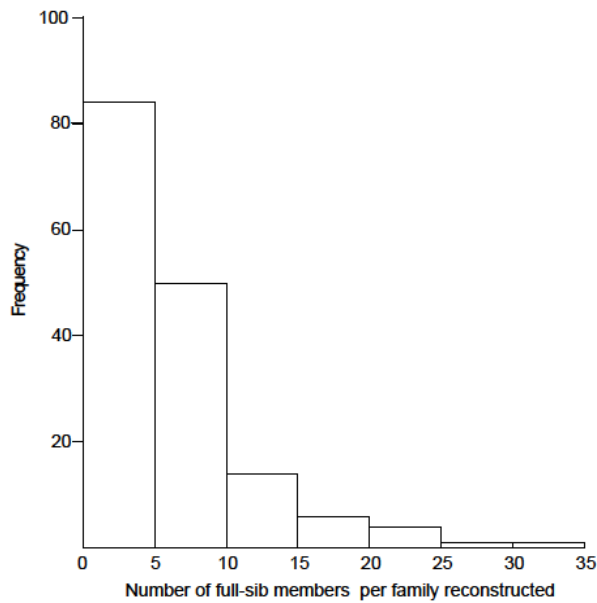

**Appendix S4.**Observed proportion of full-sibs sharing the same host withineach site (Obs) compared to the series of expected proportions of full-sibs infecting the same host withineach site under the nullhypothesis (i.e., pairs of full-sibs are distributedrandomlyamong hosts withineach site) obtainedafter 10,000 permutations of the host matrix.

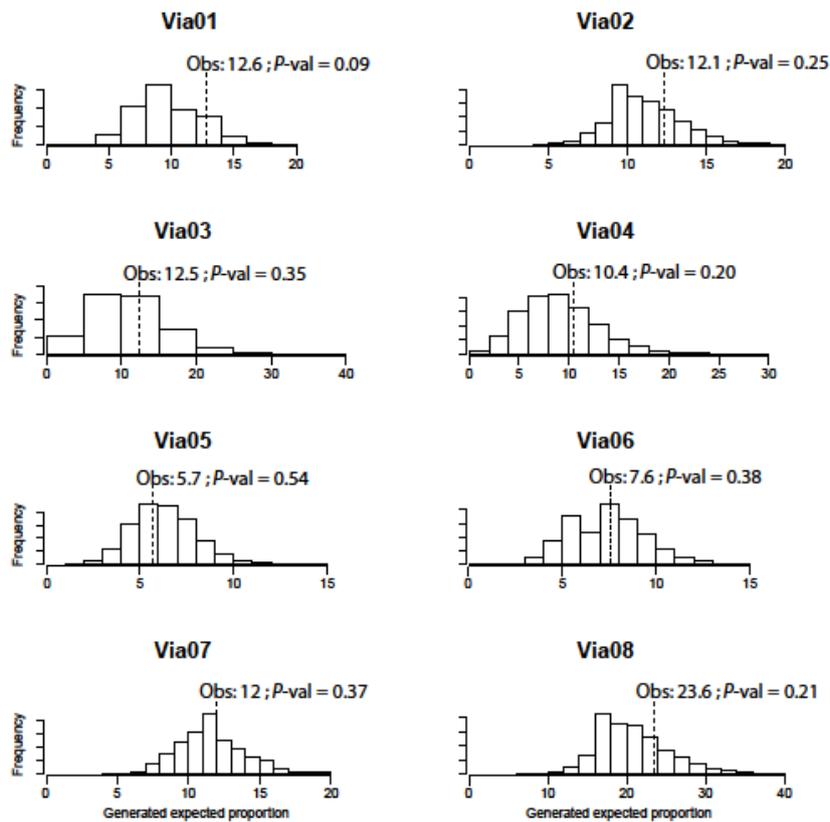

**Appendix S5.** Distribution of the family core location estimated for the reconstructed full-sib families of

*T. Polycolpus* including more than five full-sibs.

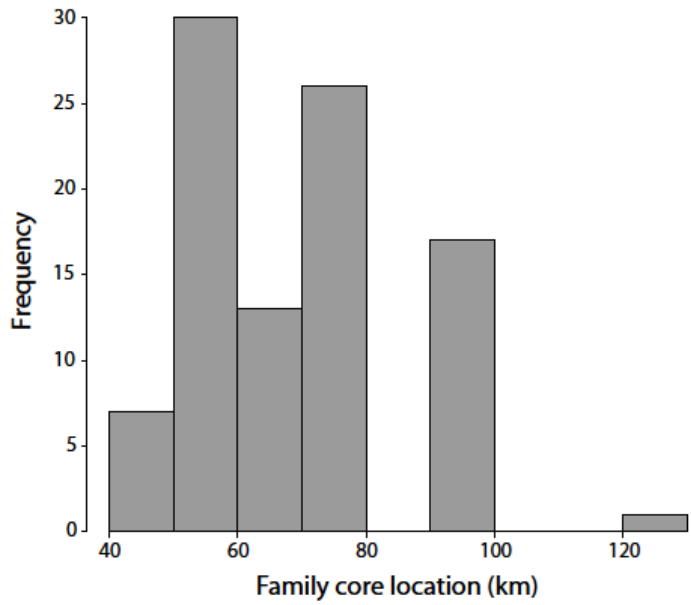